# A Novel Procedural Generation for Level Design of Mansions and Dungeons

**Isaac Fiuza Vieira**[1], **Kathya Silvia Collazos Linares**[1], **Esteban Walter Gonzalez Clua**[2], **Érick Oliveira Rodrigues**[1, 3]

[1] Departamento de Informática - Universidade Tecnológica Federal do Paraná (UTFPR)
Pato Branco, PR - Brazil

[2]Departamento de Ciência da Computação - Universidade Federal Fluminense (UFF)
Niterói, RJ - Brazil

[3]Laboratório Nacional de Computação Científica - LNCC
Petrópolis, RJ - Brazil

isaacvieira@alunos.utfpr.edu.br, kathya@utfpr.edu.br

esteban@ic.uff.br, erickr@lncc.br

***Abstract.*** ***Introduction****: Procedural Content Generation (PCG) has become an essential technique in game development due to its ability to reduce production time and cost while increasing replayability and variety. However, when not aligned with level design principles, PCG can lead to incoherent spatial structures and poor gameplay experiences.* ***Objective****: This work proposes a PCG method guided by level design principles to generate structured indoor environments — such as houses, mansions, and dungeons — aiming to ensure both architectural coherence and navigability.* ***Methodology****: The method is divided into three main stages: segmentation of the space using Binary Space Partitioning (BSP); logical connection of rooms based on graph traversal to prevent redundant links; and a post-processing stage responsible for cleaning structural artifacts and improving visual cohesion. The methodology allows parameterization of room area and shape, with randomness controlled via seeds for reproducibility.* ***Results****: Two experiments were conducted. The first demonstrated the flexibility of the methodology under different seeds and parameter configurations. The second evaluated the navigability of generated maps by verifying connectivity using Breadth-First Search (BFS). In this test, 100,000 maps were generated, and with suitable parameters, over 91% of them achieved complete connectivity.*

## 1. Introduction

The video game industry generates approximately 67 billion dollars in revenue, considering both console (hardware) and game (software) sales, surpassing other traditional media such as music, film, and literature [Marchand e Hennig-Thurau 2013]. This significant growth, observed over the past 25 years, has been driven by continuous technological advances, positioning the game industry as one of the most dynamic and influential sectors on the global stage.

Game development, however, is a complex and multidisciplinary process that involves several interdependent elements. Among the key challenges faced by developers are level design, which defines the structure and pacing of the gameplay experience; narrative design, which provides depth and emotional engagement; game mechanics, which shape interactivity and challenge; and art direction, which contributes to the game's visual appeal and immersive atmosphere. All of these factors directly influence the quality of the player experience, particularly in terms of engagement and immersion [Jennett et al. 2008].

In this context, this work proposes a procedural generation approach for indoor environments—such as mansions and dungeons—guided by level design principles. The proposed method aims to automate the creation of coherent, navigable, and diverse maps using techniques such as Binary Space Partitioning (BSP), graph-based analysis, and post-processing. The goal is to ensure that the generated environments are not only functional but also structurally designed to support gameplay and enhance player immersion.

## 2. Related Work

Procedural Content Generation (PCG) refers to the automated creation of game elements—such as levels, maps, quests, textures, characters, and objects—through algorithms. This technique has been employed since the early days of the game industry, with iconic examples such as Rogue and Elite, and remains highly relevant in modern titles like Minecraft, The Binding of Isaac, and Hades [Zhang et al. 2022].

PCG enables the efficient creation of environments, challenges, structures, and narratives, promoting greater variety and replayability while optimizing development resources [De Kegel e Haahr 2020]. Although strongly associated with game production, the technique also finds applications in other creative domains, such as the animation industry. A notable example is Pixar's RenderMan system, which uses procedural techniques to automatically generate complex materials and textures [Freiknecht e Effelsberg 2017].

In the context of games, PCG is especially valued for its ability to generate vast, unique, and dynamic worlds that adapt to each player's experience [Short e Adams 2017], while also reducing production time and development costs [van der Linden et al. 2014]. One of the most notable large-scale uses of PCG is in No Man's Sky, which employs seed-based algorithms to generate over 4 quintillion unique, explorable planets [Tait e Nelson 2022], a scale that would be impractical through manual construction.

Despite technological advances, game development remains a complex and multidisciplinary process. One of its main challenges is level design, responsible for structuring the pacing and progression of the experience. Game design can be understood as the process of defining the game's goals, functionalities, and elements. Even when adding a simple object, such as a chair, designers must consider whether it will be interactive, functional, or merely decorative — decisions that impact its role in the game world [Kremers 2009].

Beyond individual elements, game design also involves the structuring of scenes and gameplay mechanics. It defines whether a given area will focus on combat, exploration, puzzle-solving, or narrative progression. These choices shape the player's journey and directly influence NPC behavior, challenge level, and gameplay pacing [Kremers 2009].

Despite its potential, PCG may produce environments with poor structure or limited navigability when not guided by level design principles. Without clear spatial organization and player-centered considerations, generated content can lack balance, flow, and clarity [Fernandes et al. 2021]. This highlights the importance of integrating level design into procedural methods to ensure coherent and engaging experiences.

To address these limitations, recent research has focused on PCG approaches that incorporate level design principles to improve spatial coherence and navigability. The literature classifies these approaches into three main categories: traditional methods, search-based methods, and machine learning-based methods [Zhang et al. 2022]. A comparative summary of the main characteristics of these approaches is presented in Table 1. This comparison highlights how different techniques focus on aspects such as spatial structure, navigability, gameplay flow, or narrative coherence, and how the proposed method differs by integrating multiple level design principles into the procedural generation process.

**Table 1. Comparative analysis of PCG approaches.**

| Work | Techniques | Focus | Evaluation | Application |
|---|---|---|---|---|
| Putra et al. (2023) | BSP + L-System | Spatial structure, connectivity | Quantitative | Dungeon layout |
| Lima et al. (2019) | Genetic Algorithm + Automated Planning | Narrative coherence, player goals | Quantitative + Qualitative | Mission generation |
| Khalifa et al. (2019) | Visual Entropy Analysis | Spatial clarity, navigability | Quantitative | Tile-based games |
| Proposed (This Work) | BSP + Graph + Post-processing | Spatial coherence, connectivity, gameplay flow | Quantitative | Indoor environments |

Traditional methods—such as pseudorandom generators, noise functions (e.g., Perlin Noise), and formal grammars—are noted for their simplicity and efficiency, and are widely used in the generation of mazes, terrains, and dungeons. Search-based methods rely on evolutionary algorithms and evaluation functions to iteratively refine generated content, while machine learning-based methods use data-driven models to produce content adapted to existing patterns or learned player preferences.

Within the scope of constructive approaches, Putra, Tarigan, and Zamzami (2023) propose a technique that combines Binary Space Partitioning (BSP) with formal grammars, specifically L-systems, to generate corridors that connect pre-designed rooms in dungeon maps. The generation process begins with spatial division using BSP, followed by the ordered placement of rooms and the application of stochastic grammar rules to create connecting corridors. Their work focuses on the topological generation of maze-like environments and evaluates measures such as execution time and structural complexity.

While both approaches use BSP, the present work differs by dynamically generating rooms based on geometric criteria and employing graph-based logic to define connections, prioritizing structural coherence and navigability. Whereas Putra et al. emphasize formal variation through grammars, this study seeks to integrate procedural generation with level design principles, ensuring adaptable and functionally explorable environments.

Complementing this perspective, Lima et al. (2019) propose a procedural mission generation system based on the combination of genetic algorithms and automated planning techniques. Their approach considers factors such as player progression, narrative context, and storyline coherence, allowing for the creation of dynamic, adaptive missions. The study also presents a technical performance evaluation—comparing basic, parallelized, and optimized versions of the model—and includes user testing to assess player perceptions of procedurally generated missions versus those created manually by designers. However, the focus remains on player actions and objectives, without addressing the spatial organization of the environments where these actions occur.

On the other hand, the importance of structural coherence in PCG is emphasized in the work of [Khalifa et al. 2019], which analyzes the relationship between visual entropy and navigability in tile-based games. The article shows that high-entropy levels—characterized by abrupt and disorganized visual changes—impair player clarity and progression. The study proposes measures such as horizontal consistency to ensure that scene elements are arranged logically and intuitively, reinforcing the need for design guidelines even in automated generation processes.

While Putra et al. (2023) explore grammar-based dungeon layout construction and Lima et al. (2019) address emergent narrative in adaptive missions, Khalifa et al. (2019) highlight that effective PCG depends on the articulation between randomness and intentional design. This triad—topology, narrative, and spatial coherence—underscores the need for approaches that not only generate content but do so in alignment with fundamental game design principles.

Given this context, a significant gap in the literature becomes evident: the lack of methods focused on the procedural generation of structured indoor environments—such as houses, mansions, and dungeons—that balance structural variety with navigability, architectural coherence, and alignment with level design. The present work aims to address this gap by proposing an approach that integrates procedural generation with heuristics for accessibility and spatial organization, ensuring not only functional results but also fluid and immersive player experiences.

## 3. Methodology

The procedural generation process was structured into three main stages: the division of the initial space using an algorithm based on Binary Space Partitioning (BSP), the connection of rooms using a graph-based logic, and finally, a post-processing step responsible for fine-tuning the generated structure. This process was guided by core level design principles aimed at ensuring not only functional map generation but also a meaningful gameplay experience. Specifically, the method prioritizes spatial coherence (clear room and corridor organization), navigability (ensuring players can traverse the environment efficiently without dead ends or inaccessible areas), and gameplay flow (supporting a logical progression of exploration and challenge). Figure 1 illustrates the flow of these stages, as well as the processes involved in each phase.

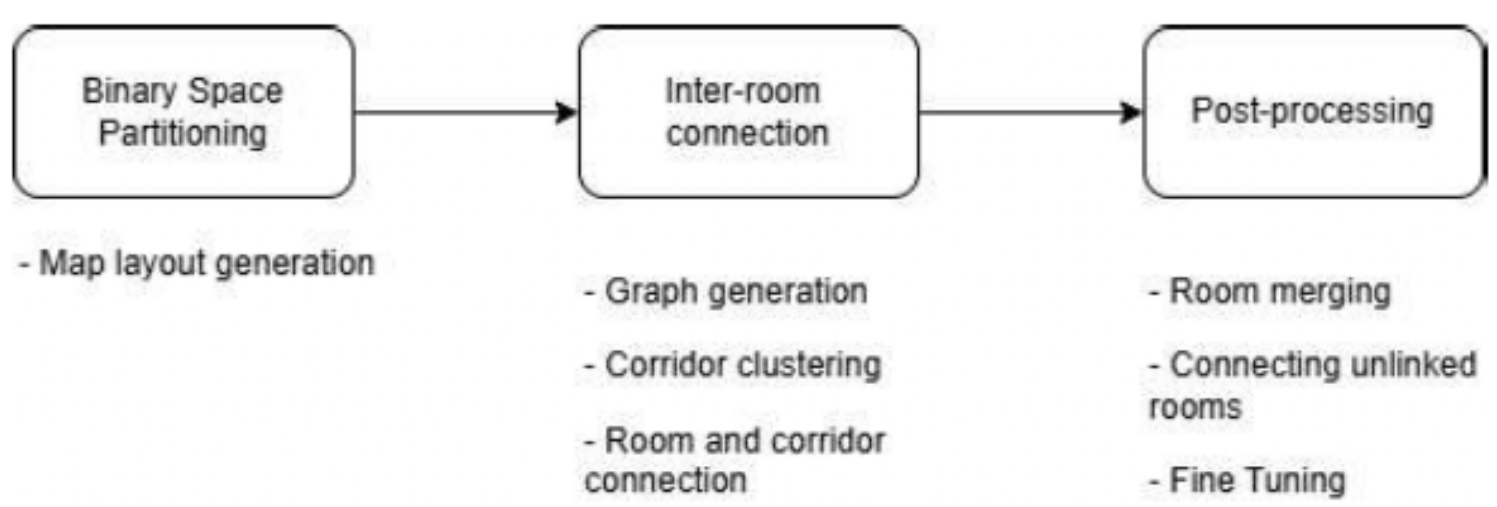

**Figure 1. Stages of the procedural generation process and the operations involved in each phase.**

The Binary Space Partitioning (BSP) technique consists of recursively dividing a space into smaller subregions based on predefined criteria [Putra et al. 2023]. This approach is widely used in several areas of computing, including procedural map generation, spatial indexing, collision detection, robotics, and graphical rendering [Fan et al. 2018]. The partitions can follow either horizontal or vertical orientations, determined by parameters such as minimum dimensions, specific priorities, or predefined patterns.

In the first stage of the procedural map generation, the BSP technique was implemented using a custom algorithm designed to operate over a matrix-based grid representation of the map. The algorithm recursively partitions the space into subregions, defining room boundaries and structural layout according to the generation parameters.

The choice of BSP is motivated by its ability to produce geometrically regular and well-defined divisions, which is particularly suitable for indoor environments such as mansions or dungeons. These types of spaces often feature rectangular layouts, linear corridors, and rooms with structured shapes. This characteristic makes the technique ideal for building architecturally coherent environments in games with such settings.

The algorithm performs successive recursive divisions of the matrix, evaluating at each iteration whether the current subarea meets two main criteria: a minimum area required to be considered a valid room (*minRoomArea*), and a minimum side length (*minRoomSide*). If both conditions are satisfied, the region is registered as a valid room; otherwise, it is subdivided into two new regions.

Similar to this approach, Putra et al. (2023) also define control parameters to guide the procedural generation process. In their method, user-defined rules allow control over corridor generation via L-Systems, including parameters such as maximum iteration limits, corridor dimensions, and permitted turning angles. These constraints serve a similar purpose — enabling structured variation and user control over the generated content — reinforcing the importance of parameterization in balancing randomness and design intent.

The choice of division direction—vertical or horizontal—is controlled by a boolean variable that alternates at each recursive call, ensuring a more balanced distribution. The auxiliary function *split* performs the division, randomly selecting a valid position to draw the dividing line based on a given seed. The use of a seed ensures the reproducibility of the process, allowing the generation of maps with identical structure when required.

The resulting segmentation classifies the generated regions as either rooms or corridors, based on their geometry. The *minRoomSide* parameter acts as a filter: regions with at least one side smaller than this threshold are considered corridors, while those with more balanced proportions are classified as rooms. Conversely, the *minRoomArea* parameter imposes an upper size limit: regions larger than this threshold are recursively subdivided to avoid the creation of excessively large rooms.

The full logic is implemented in function Algorithm 1, which combines deterministic and random elements to ensure both structural variety and consistency in the generated maps.

**Algorithm 1** Layout generation algorithm using proposed BSP.

**Require:** $startX, startY, endX, endY, toggle$
**Ensure:** Rooms and walls carved in grid $M$
1: **procedure** `Generate`$(startX, startY, endX, endY, toggle)$
2: **if** $(endX - startX) \cdot (endY - startY) < minRoomArea$ **or** $(endX - startX < minRoomSide)$ **or** $(endY - startY < minRoomSide)$ **then**
3: Create room from $(startX, startY)$ to $(endX, endY)$
4: Mark corresponding cells in $M$ as 1 {Floor}
5: **else if** $toggle$ is true **then**
6: `Split`$(startX, endX, startY, endY$, true, false$)$
7: **else**
8: `Split`$(startX, endX, startY, endY$, false, true$)$
9: **end if**
10: **end procedure**
11: **procedure** `Split`$(startX, endX, startY, endY, isVertical, toggle)$
12: **if** $isVertical$ **then**
13: $splitPosition \leftarrow$ random integer between $(startX + minRoomSide/2)$ and $(endX - minRoomSide/2)$
14: **for** $y \leftarrow startY$ **to** $endY - 1$ **do**
15: $M[splitPosition][y] \leftarrow 0$ {Wall}
16: **end for**
17: `Generate`$(startX, startY, splitPosition, endY, toggle)$
18: `Generate`$(splitPosition, startY, endX, endY, toggle)$
19: **else**
20: $splitPosition \leftarrow$ random integer between $(startY + minRoomSide/2)$ and $(endY - minRoomSide/2)$
21: **for** $x \leftarrow startX$ **to** $endX - 1$ **do**
22: $M[x][splitPosition] \leftarrow 0$ {Wall}
23: **end for**
24: `Generate`$(startX, startY, endX, splitPosition, toggle)$
25: `Generate`$(startX, splitPosition, endX, endY, toggle)$
26: **end if**
27: **end procedure**

To establish room connectivity, a logic inspired by conventional residential architecture was adopted, in which corridors function as circulation axes interconnecting the various spaces. Accordingly, each room must have at least one doorway facing an adjacent corridor. However, when this heuristic is applied directly, procedural generation may result in an excessive number of redundant connections between rooms and corridors—that is, a single room may be connected multiple times to the same corridor. This behavior compromises the structural coherence of the map, as illustrated in Figure 2.

To mitigate this issue, two complementary strategies were implemented. The first involved building a graph where each node represents a room (either a bedroom or corridor), and each edge denotes adjacency between two rooms. Based on this graph, a breadth-first search (BFS) traversal was applied to identify interconnected corridor groups—such as Group A, composed of corridors 1, 2, and 3, and Group B, composed of corridors 4 and 5—as shown in Figure 3. In this figure, the corridors are grouped into two distinct sets, outlined by dotted circles. Based on this segmentation, each room is configured to connect to at most one corridor from each group, reducing redundant links and resulting in a more logical and efficient map structure.

**Figure 2. Visual example of a map with redundant and excessive room connections.**

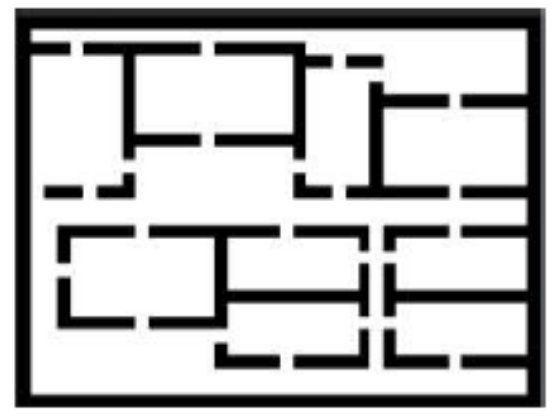

This grouping process is based on a list of rooms and an adjacency matrix, and traverses the corridors to identify connected regions. Each corridor is assigned to a group according to its connectivity with adjacent corridors, using BFS to mark all elements belonging to the same cluster. As a result, a vector is generated associating each corridor with its corresponding group, enabling the connection logic between rooms and corridors to be managed more coherently.

The process begins by iterating over the room list. If a room has already been assigned to a group or is not a corridor, it is skipped. Otherwise, the group identifier is incremented, and a new BFS is initiated from that room.

During the BFS execution, all corridors directly connected to the initial room are visited and assigned to the current group. This ensures that all interconnected corridors are grouped consistently. The search continues until there are no unvisited corridors remaining in the current cluster.

**Figure 3. Visual example of corridor group formation outlined by the dotted line.**

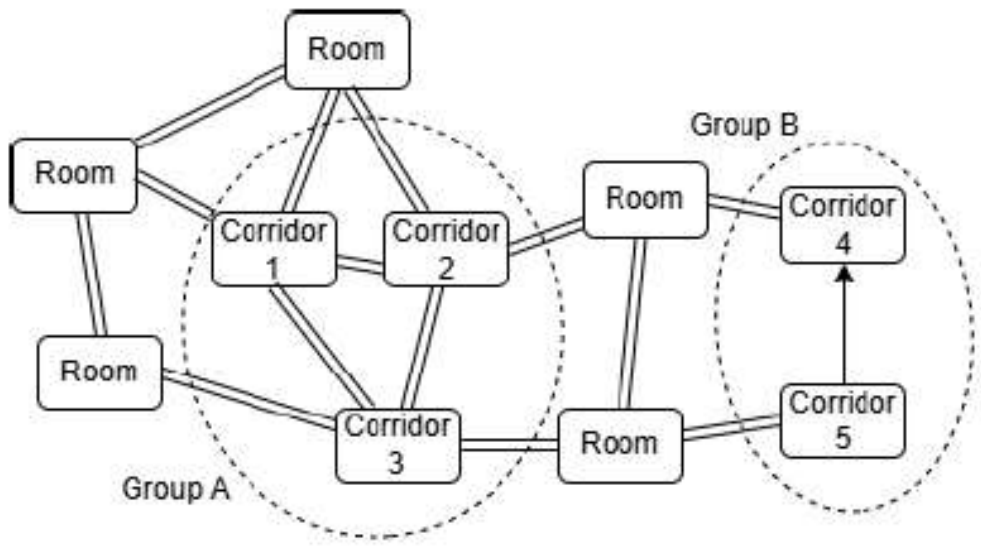

In the post-processing stage, two main actions are performed with the aim of enhancing both visual cohesion and navigability of the generated map. The first involves removing walls between corridors that belong to the same group, as identified in the previous step. This operation results in wider and more interconnected spaces, reflecting a more organic and realistic architectural structure. Additionally, a fine-tuning step is performed on the generated grid to correct small structural artifacts—such as isolated or misaligned walls—produced during segmentation.

Finally, Figure 4 presents a complete visual example of a 20x30 map generated using seed 7451143 with parameters *minRoomArea* = 100 and *minRoomSide* = 5, highlighting each stage of the process: space segmentation, room connection definition, and post-processing application.

**Figure 4. Visual example of map generation.**

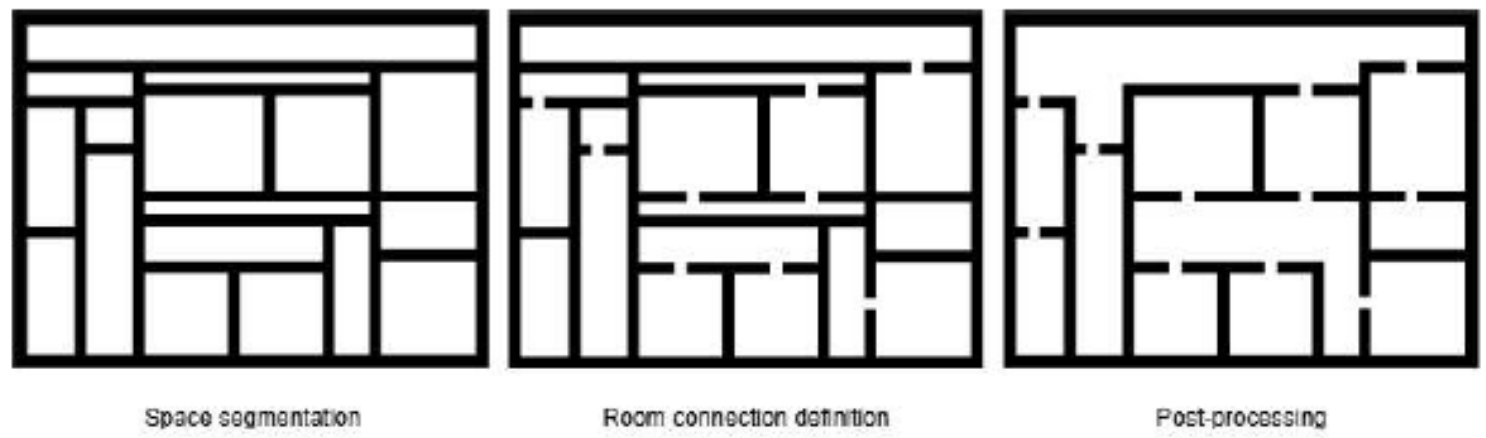

To validate room connectivity, a breadth-first search (BFS) algorithm was applied across the entire grid to verify whether all rooms were accessible from a single starting point. This approach enabled the evaluation of a large number of generated maps and the inference of the proportion that achieved full connectivity.

Inspired by the work of Khalifa et al. (2020), which proposes systematic methods for validating the playability of procedurally generated levels, our BFS-based verification aligns with the same intent of ensuring structural quality. While Procedural Content Generation via Reinforcement Learning (PCGRL) uses reinforcement learning to induce the creation of playable levels from the outset, our approach acts as a post-processing validation filter, discarding disconnected maps. In both cases, the emphasis is placed on playability and spatial accessibility.

Furthermore, just as PCGRL leverages one-hot encodings and spatial transformations to assist agents in interpreting their environment, our BFS implementation operates on a discretized map representation, ensuring efficiency in the analysis process.

## 4. Results

This section presents the results obtained from the implementation of the proposed method, organized into two main parts. First, a collection of maps generated using different seeds and parameter configurations is shown, with the aim of demonstrating the flexibility and variability of the methodology. Next, a quantitative experiment is described to evaluate the connectivity of the generated maps by analyzing the proportion of cases in which all rooms are accessible, thereby ensuring full navigability of the environment.

To illustrate the system's ability to generate varied layouts, maps were produced using different seeds and combinations of parameters, such as the minimum room area and the minimum side length. This analysis aims to highlight the method's adaptability to different constraints while maintaining structural coherence and functional room arrangements.

Figure 5 presents the results of a visual experiment involving controlled variation of the generation parameters. In the first column, the *minRoomSide* value was fixed at 5, while the *minRoomArea* varied between 80 and 120. In the second column, the minimum area was fixed at 100, and the *minRoomSide* ranged from 3 to 7. The examples show how these variations directly affect the complexity, segmentation, and architectural patterns of the mansion layouts.

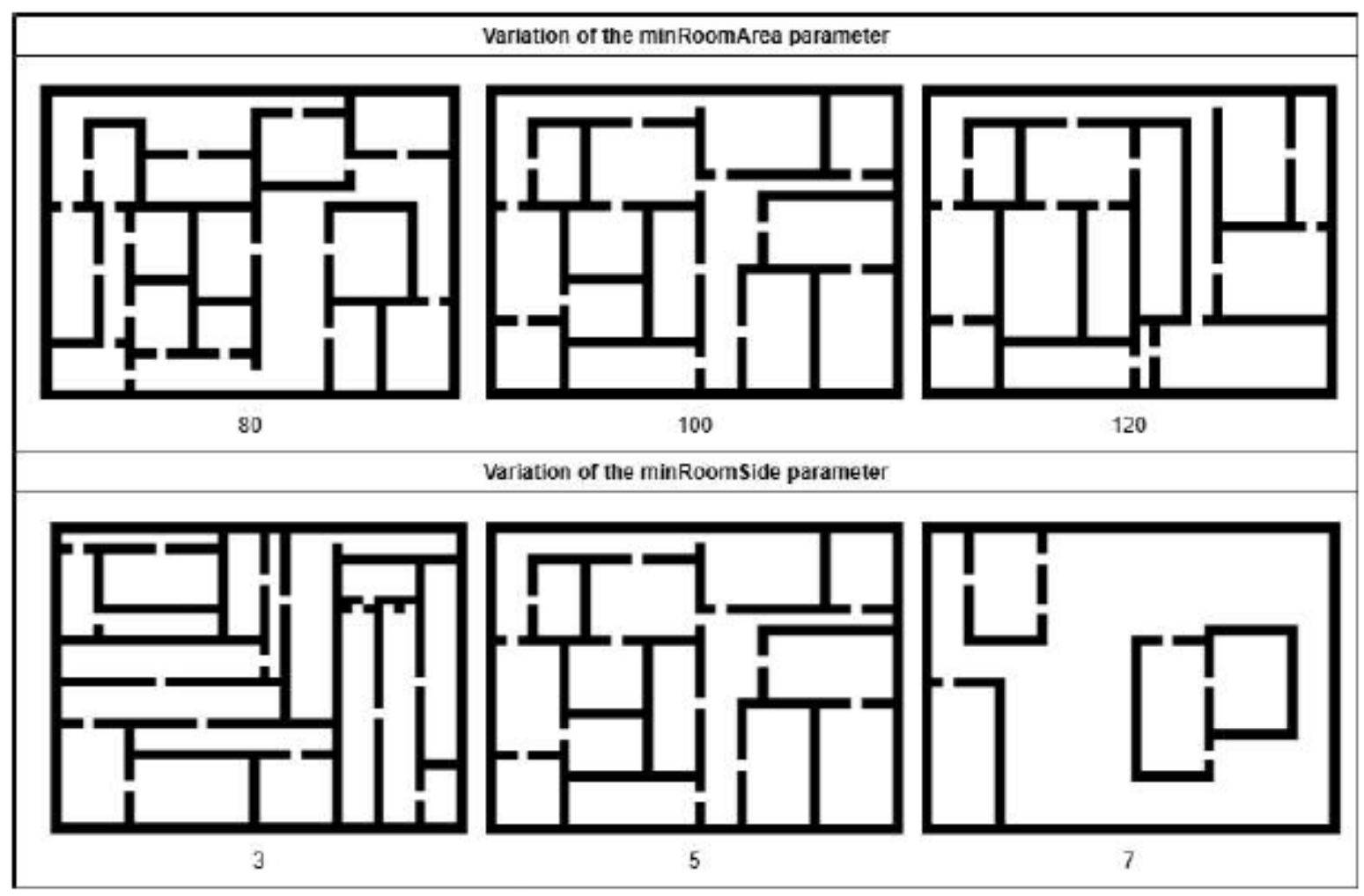

**Figure 5. Generated layouts with varied input parameters.**

It is observed that the diversity of the layouts strongly depends on the choice of input parameters. Thus, for each map size, it is necessary to adjust and empirically test the generation values in order to achieve a balance between structural variety and spatial coherence in the final layout.

In addition to the quantitative evaluation, a qualitative comparison with the approach by Putra et al. (2023) is shown in Figure 6. While both use BSP for space partitioning, Putra's method generates isolated rooms connected by long corridors, resulting in low adjacency. In contrast, the proposed method creates a denser layout, with interconnected rooms organized around central corridors, reducing navigational cost and enhancing immersion by better reflecting real-world architectural logic. These results highlight the potential of the proposed method to produce more coherent layouts aligned with gameplay flow.

Regarding connectivity, two experiments were conducted, each involving 100,000 procedurally generated maps with dimensions of 30x40. Connectivity verification was performed as described in the methodology section, assessing whether all rooms were accessible from a single initial point.

In the first scenario, using more restrictive parameters (*minRoomArea* = 100 and *minRoomSide* = 5), approximately 85.9% of the maps exhibited full connectivity. In the second scenario, where the minimum room area was reduced to 80 while maintaining the same *minRoomSide*, the connectivity rate increased to 91.6%.

These results suggest that less restrictive parameters in the space segmentation process contribute to greater interconnectivity between rooms, thereby facilitating the creation of fully accessible layouts. This characteristic is crucial for ensuring spatial coherence and viable navigation in procedurally generated maps.

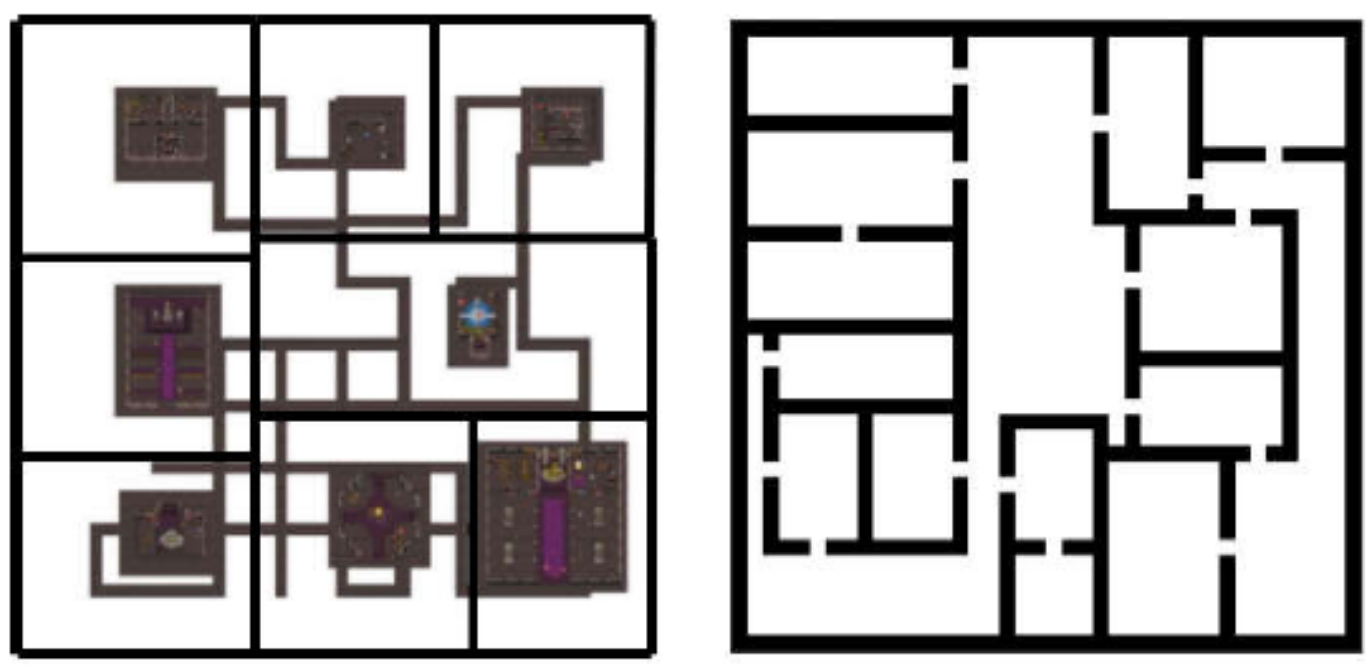

**Figure 6. Comparison between the dungeon layout generated by Putra et al. (2023) (left, adapted for comparative purposes) and the proposed method (right).**

The source code and a demonstration GIF illustrating the generation process are available on GitHub [Vieira 2025] to support reproducibility and further research.

## 5. Conclusion

This work proposed a procedural content generation approach guided by level design principles for the creation of indoor environments in digital games, such as houses, mansions, and dungeons. Unlike traditional approaches that focus solely on generating random layouts, the presented solution integrates Binary Space Partitioning (BSP), graph-based modeling, and post-processing techniques, emphasizing architectural coherence, navigability, and a fluid player experience.

The distinctive aspect of this approach lies in the explicit incorporation of connectivity heuristics and spatial structuring aligned with design intent. By considering factors such as room accessibility and internal layout logic, the methodology not only produces varied maps but also fosters playable, cohesive environments adaptable to different gameplay styles.

The conducted experiments demonstrated the methodology's flexibility under parameter variation and confirmed its effectiveness in producing functionally connected layouts - achieving up to 91.6% of fully connected maps out of 100,000 generated samples. These results highlight the potential of procedural generation as a valuable tool for supporting game development, especially in scenarios with limited resources or high demands for replayability.

To further enhance this approach, future improvements may focus on strategies to guarantee full connectivity in all cases, automate parameter tuning for greater adaptability, and incorporate functional semantics to define room purposes and gameplay pacing. Moreover, complementing the structural evaluation with user-centered assessments could provide valuable insights into player experience and spatial perception.

## References

De Kegel, B. e Haahr, M. (2020). Procedural puzzle generation: A survey. *IEEE Transactions on Games*, 12(1):21–40.

de Lima, E. S., Feijó, B., e Furtado, A. L. (2019). Procedural generation of quests for games using genetic algorithms and automated planning. In *SBGames*, pages 144–153.

Fan, X., Li, B., e Sisson, S. (2018). The binary space partitioning-tree process. In *International Conference on Artificial Intelligence and Statistics*, pages 1859–1867. PMLR.

Fernandes, P. M., Jørgensen, J., e Poldervaart, N. N. T. G. (2021). Adapting procedural content generation to player personas through evolution. In *2021 IEEE Symposium Series on Computational Intelligence (SSCI)*, pages 01–09.

Freiknecht, J. e Effelsberg, W. (2017). A survey on the procedural generation of virtual worlds. *Multimodal Technologies and Interaction*, 1(4).

Jennett, C., Cox, A. L., Cairns, P., Dhoparee, S., Epps, A., Tijs, T., e Walton, A. (2008). Measuring and defining the experience of immersion in games. *International Journal of Human-Computer Studies*, 66(9):641–661.

Khalifa, A., Bontrager, P., Earle, S., e Togelius, J. (2020). Pcgrl: Procedural content generation via reinforcement learning. In *Proceedings of the AAAI Conference on Artificial Intelligence and Interactive Digital Entertainment*, volume 16, pages 95–101.

Khalifa, A., Green, M. C., Barros, G., e Togelius, J. (2019). Intentional computational level design. In *Proceedings of the Genetic and Evolutionary Computation Conference*, GECCO '19, page 796–803, New York, NY, USA. Association for Computing Machinery.

Kremers, R. (2009). *Level design: concept, theory, and practice*. CRC Press.

Marchand, A. e Hennig-Thurau, T. (2013). Criação de valor na indústria de videogames: Economia da indústria, benefícios ao consumidor e oportunidades de pesquisa. *Jornal de Marketing Interativo*, 27(3):141–157.

Putra, P. A., Tarigan, J. T., e Zamzami, E. M. (2023). Procedural 2d dungeon generation using binary space partition algorithm and l-systems. In *2023 International Conference on Computer, Control, Informatics and its Applications (IC3INA)*, pages 365–369.

Short, T. e Adams, T. (2017). *Procedural generation in game design*. CRC Press.

Tait, E. R. e Nelson, I. L. (2022). Nonscalability and generating digital outer space natures in no man's sky. *Environment and Planning E*, 5(2):694–718.

van der Linden, R., Lopes, R., e Bidarra, R. (2014). Procedural generation of dungeons. *IEEE Transactions on Computational Intelligence and AI in Games*, 6(1):78–89.

Vieira, I. (2025). Bsp-procedural-generation. `https://github.com/isaacfvi/BSP-Procedural-generation`. Accessed: June 2025.

Zhang, Y., Zhang, G., e Huang, X. (2022). A survey of procedural content generation for games. In *2022 International Conference on Culture-Oriented Science and Technology (CoST)*, pages 186–190.